\documentclass[%
 reprint,
showpacs,preprintnumbers,
 amsmath,amssymb,
 aps,
pra,
]{revtex4-1}

\usepackage{graphicx}
\usepackage{dcolumn}
\usepackage{bm}
\usepackage{epstopdf}

\begin{document}


\title{Dipolar dissociation dynamics in electron collisions with oxygen molecules}

\author{Pamir Nag}
\email{pamir1118@iiserkol.ac.in}
\author{Dhananjay Nandi}%
 \email{dhananjay@iiserkol.ac.in}
\affiliation{Indian Institute of Science Education and Research Kolkata, Mohanpur 741246, India
}%




\date{\today}

\begin{abstract}

The dipolar dissociation of molecular oxygen due to 21-35 eV energy electron collision has been studied using the time sliced velocity map imaging technique. A rough estimation about the threshold of the process and the kinetic energy and angular distribution of the fragment negative ions are measured. The dipolar dissociation found to be occur due to pre-dissociation of a Rydberg state via ion-pair state for lower incident electron energies as well from also direct excitation to the ion-pair states for relatively higher primary beam energy. The location and symmetry of the excited states were determined from the kinetic energy and angular distribution data respectively.

\end{abstract}

\maketitle


\section{Introduction}
Dissociative ionization by electron impact plays an important role in astrochemistry and biology. In dipolar dissociation the molecule dissociates into an anionic and cationic fragments. The dipolar dissociation process starts around above 15 eV incident electron energy. Unlike the well studied dissociative electron attachment (DEA) process \cite{o2:van_DEA, o2:dn_cross} in dipolar dissociation the incoming electron does not resonantly get captured by the molecule but, might partially transfers its kinetic energy to the molecule and leave it in an excited state. The molecules populated into an ion-pair state either via pre-dissociation of a Rydberg state or through direct excitation dissociate into  a cationic and anionic fragments.

\begin{equation}
\text{O}_2 + e^- \rightarrow \text{O}^* _2 +e^- \rightarrow \text{O}^+ + \text{O}^- +e^-
\end{equation}

The ion-pair dissociation dynamics of O$_2$ is well studied mainly around the threshold using lasers \cite{o2:baklanov, o2:zhou, o2:chupka}. Primarily the pre-dissociation of a Rydberg state into an ion-pair state was found to be responsible for the dipolar dissociation process around the threshold. But only a few studies of the dipolar dissociation well above the threshold were performed. It is difficult to access the ion-pair states using lasers due to the required energy range. But the same ion-pair states can easily be probed using electron beam of energy above around 15 eV. Van Brunt and Kieffer \cite{o2:polar_van1} studied the kinetic energy and angular distribution of the O$^-$ ions produced through dipolar dissociation over a limited angular range back in 1974. To best of our knowledge no recent study except by Nandi \emph{et al.} \cite{o2:polar_DN} is available in literature. In that article also the authors only reported the data but did not perform detailed analysis of the ion-pair dissociation process. In the present article the dipolar dissociation dynamics of molecular oxygen due to interaction with electrons having energies in between 21 to 35 eV is studied using well established velocity slice imaging technique \cite{inst:DN,inst:adaniya,inst:moradmand}. In the present studied only the O$^-$ fragments were probed and it was assumed that a O$^+$ ion is always accompanied with the anionic fragments as negative ions can only be formed due to dipolar dissociation processes \cite{lozier, dorman} around the reported energy region. From the velocity slice images the kinetic energy and angular distribution of the O$^-$ ions were obtained for six different incident electron energies. The location and symmetry of the ion-pair states involved in the process were determined from those measurements.

\section{Experiment}
\begin{figure}
\begin{center}
\includegraphics[scale=.5]{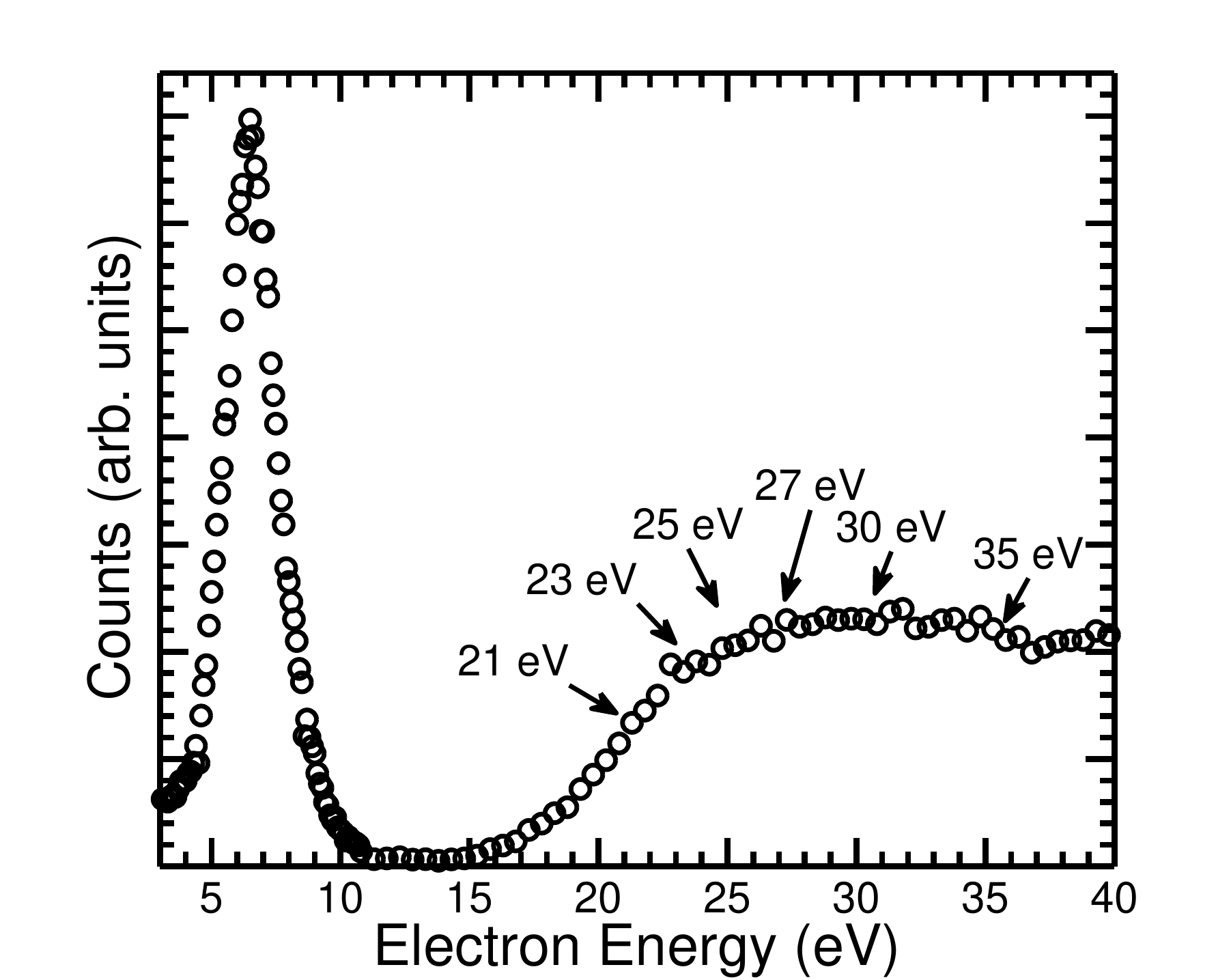}
\caption{Ion Yield curve of O$^-$ ions produced from O$_2$ due to interaction with low-energy electrons.} \label{fig:ion_yield}
\end{center}
\end{figure}
To study the dipolar dissociation dynamics in O$_2$ a time sliced velocity map imaging (VMI) spectrometer similar with previous report by Nandi \emph{et al.} \cite{inst:DN} with some minor modifications was used. The same apparatus have already been employed for the dissociative electron attachment (DEA) study to different molecules \cite{my:CO-DEA, my:co2, my:cl2} in recent time. Some details of the current apparatus can be found elsewhere \cite{my:instrum}. For sake of completeness the setup is described here in brief. 

The entire experiment was performed under oil-free ultra high vacuum condition with a base pressure of the order $\sim$ 10$^{-9}$ mbar. The chamber was heated at 130$^\circ$C for several days before performing the current experiments to remove water vapours and other impurities that might present inside the chamber wall.  A magnetically well collimated pulsed electron beam of 200 ns duration and 10 kHz repetition and of controllable energy was employed in the experiment. A custom build electron gun consists of thermally heated filament with typical energy resolution 0.8 eV was used for the purpose. A pair of magnetic coils in Helmholtz configuration, producing around 40 G uniform magnetic field, was mounted outside the chamber to collimate the low-energy otherwise diverging electron beam. A Faraday cup was placed opposite to the electron gun to monitor the time averaged pulsed electron beam. An effusive molecular beam produced from a capillary tube of 1 mm diameter was made to interact perpendicularly with the pulsed electron beam in the interaction region of the VMI spectrometer. The molecular beam was along the spectrometer axis and directed towards the detector. The VMI spectrometer is a three field time-of-flight (TOF) spectrometer \cite{inst:DN} which focuses ions starting from a finite volume onto a two-dimensional position sensitive detector (2D-PSD) such that ions of given velocity are mapped onto a single point on the detector irrespective of their spatial location in the source region. The ions produced in the interaction region due to the electron-molecule collisions were pulse extracted using a moderate electric field pulse of 4 $\mu$s duration. The extraction pulse was applied 100 ns after the electron beam pulse had passed. The delayed extraction pulse prevents the electrons from reaching the detector and also provides an appropriate time spread to the ions to expand for better time slicing. The 2D-PSD used in the experiment consists of three micro channel plates (MCP) in Z-stack configuration and a three layer delay line hexanode \cite{hex1}. The TOF of the detected ions had been determined from the back MCP signal whereas the x and y positions of each detected ions were calculated from the three anode layer \cite{hex1}. The x, y position along with TOF of each detected particles were acquired and stored in List-Mode Format (LMF) using CoboldPC software from RoentDek. The entire kinetic energy and angular distribution information can be extracted from the central slice through the `Newton Sphere' in the plane of the detector and containing the electron beam axis. The central time sliced images were obtained by selecting only the ions produced within an appropriate time window during off-line analysis of the LMF file using the same CoboldPC software.

The typical full width at half maxima (FWHM) of the TOF of the O$^-$ ions produced in the current report was about 500 ns. A 50-ns thin time sliced images around the central part of the `Newton Sphere' has been considered to get the kinematically complete information of the ions. The electron beam energy calibration was performed using the 6.5 eV DEA peak of O$^-$/O$_2$ \cite{ref:rapp}. To determine the kinetic energy of the ions from the sliced images the spectrometer was calibrated with the energy release of O$^-$/O$_2$ in DEA range \cite{o2:dn_cross} and cross-checked with the energy release of O$^-$/CO$_2$ at 8.2 eV  \cite{co2:moradmand2, my:co2}.

To get the ion yield curve a different set of data acquisition system was used. For this purpose only the signal from the MCP was taken. The MCP signal was amplified through a Fast Amp and then fed to a constant fraction discriminator (CFD). The output from CFD was connected to STOP input of a nuclear instrumentation module (NIM) standard time to amplitude converter (TAC) and START pulse was generated from a master pulse generator controlling the timing sequences of the entire experiment. The output of TAC was connected to a multichannel analyser (MCA, Ortec model ASPEC-927) and finally communicated with the data acquisition system installed in a dedicated computer via high-speed USB 2.0 (Universal Serial Bus) interface. A LabVIEW based data acquisition system was used to obtain the ion yield curve and control most of the instruments employed in the experiment. A complete details of this system can be found elsewhere \cite{my:instrum}.

\begin{figure}
\begin{center}
\includegraphics[scale=.5]{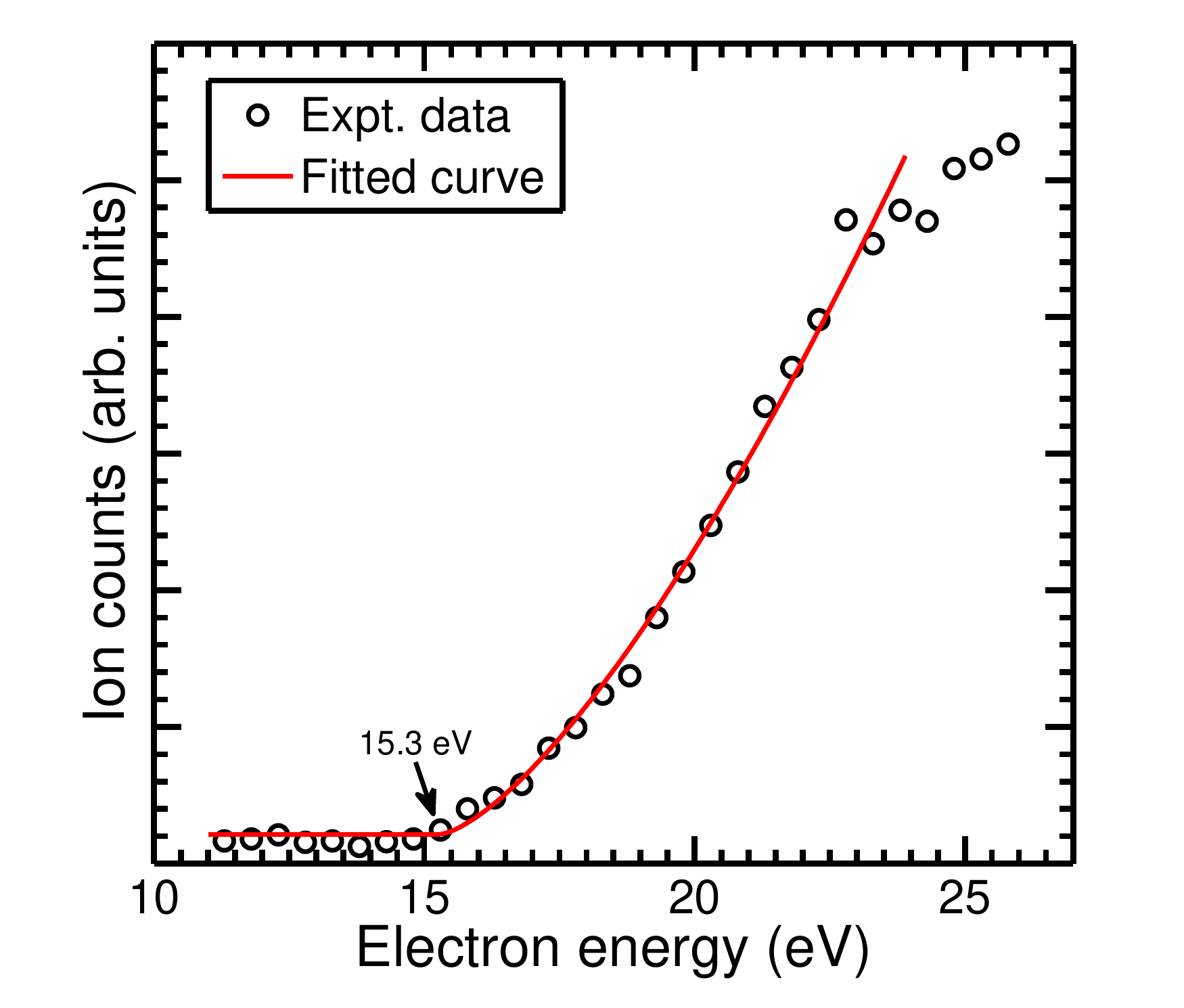} 
\caption{Ion yield curve of the O$^-$ ions produced due to dipolar dissociation around the threshold value. The fitted curve suing the couple of equation~\ref{eq:th_fit} is also shown by a solid line.} \label{fig:threshold}
\end{center}
\end{figure}

\begin{figure*}
\begin{center}
\includegraphics[scale=0.35]{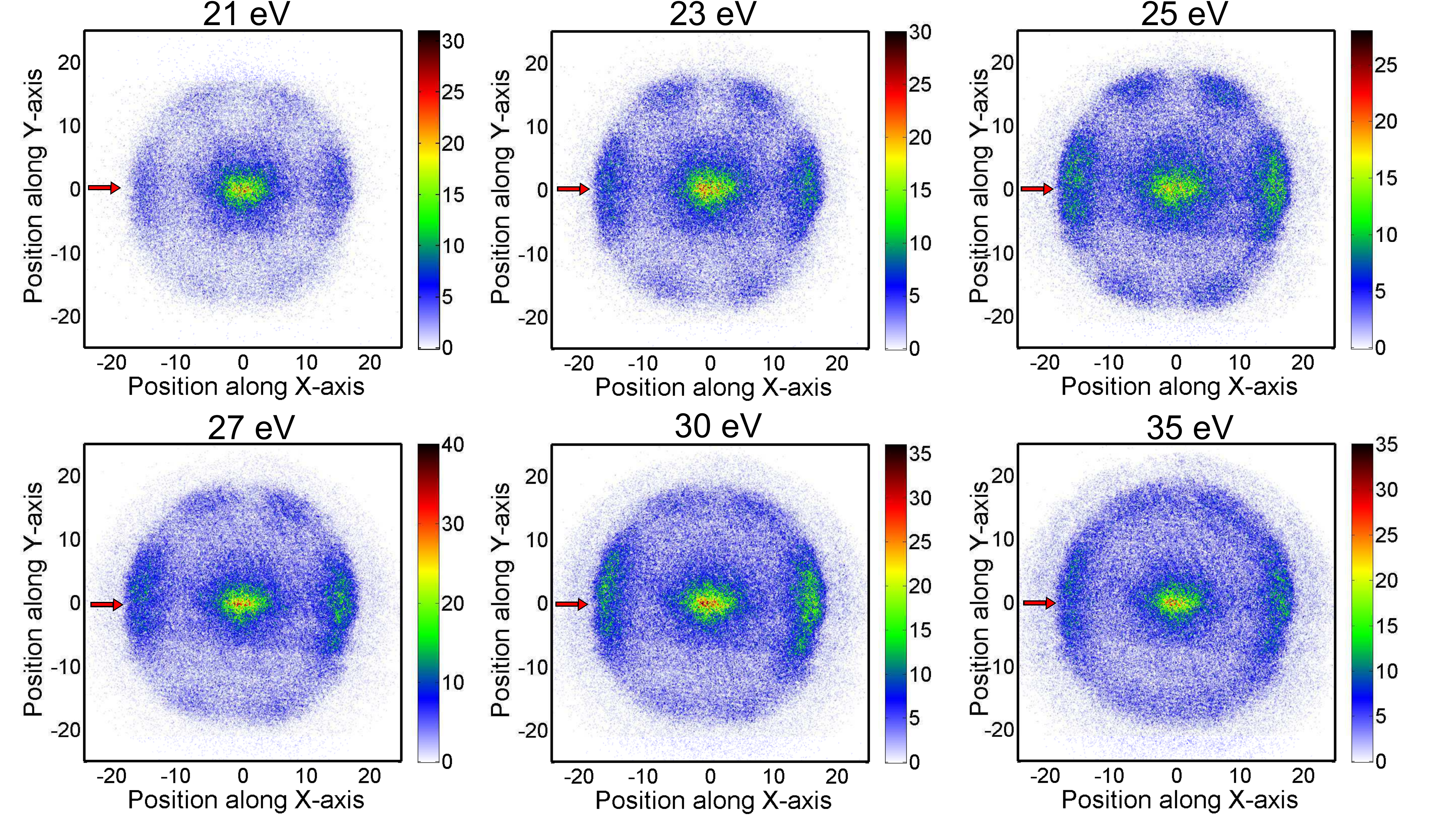}
\caption{Time sliced velocity map images of the O$^-$ ions created due to six different incident electron energies as mentioned on the top of each image. The incident electron beam direction is indicated by small arrows along the x-axis.} \label{fig:VSI}
\end{center}
\end{figure*}

\section{Result and Discussion}

The experiments have been performed using 99.9\% pure commercially available oxygen gas. The ion yield curve of the O$^-$ ions produced due to low-energy electron collision with O$_2$ molecule is shown in Figure~\ref{fig:ion_yield}. A resonant peak due to dissociative electron attachment (DEA) at 6.5 eV and almost constant ion formation due to dipolar dissociation after nearly 15.3 eV can be observed in the ion yield curve. The curve is in good agreement with the previous report \cite{ref:rapp}. The main focus of this article is to understand the dipolar dissociation dynamics starting from around 15.3 eV. The arrows in Figure~\ref{fig:ion_yield} indicates the energies at which the velocity slice images (VSI) were taken. From the ion yield curve the appearance energy for the dipolar dissociation processes has been computed whereas from the VSI the kinetic energy and angular distribution of the O$^-$ ions have been measured.

Unlike the DEA in dipolar dissociation process the incoming electrons do not get attached with the molecule but might partially transfer its kinetic energy. In the present case if $V_i$ is the amount of energy transfer to the molecule, $D=5.15$ eV \cite{web:nist} is the bond dissociation energy of O$_2$, $IP= 13.6$ and $A=1.5$ eV \cite{o2:polar_van1,o2:chantry} are the ionization potential and electron affinity of neutral O atom and $E$ is the kinetic energy of each of the O$^+$ and O$^-$ ion fragments formed in the process, then from conservation of energy one can found

\begin{equation}
V_e = (E_i +D -A + IP) + 2E
\end{equation}

The threshold energy for the diploar dissociation process using above mentioned accepted thermochemical values of different parameters and considering the O$^+$ fragments formed in ground state ($E_i=0$) is found to be 17.25 eV. From the ion yield curve it is possible to  experimentally determine the energy threshold of the DD process. The data points in the ion yield curve near the threshold is fitted using the couple of equation given by Fiegele \emph{et al.} \cite{th:fiegele} and recently used by {Szyma{\'n}ska} \emph{et al.} \cite{dd:szymanska}
\begin{eqnarray}
f(E) &=& b  \text{ for }  E<E_{Th} \nonumber \\
f(E) &=& b+a(E-E_{Th})^n \text{ for } E>E_{Th} \label{eq:th_fit}
\end{eqnarray}

\begin{figure}
\begin{center}
\includegraphics[scale=0.49]{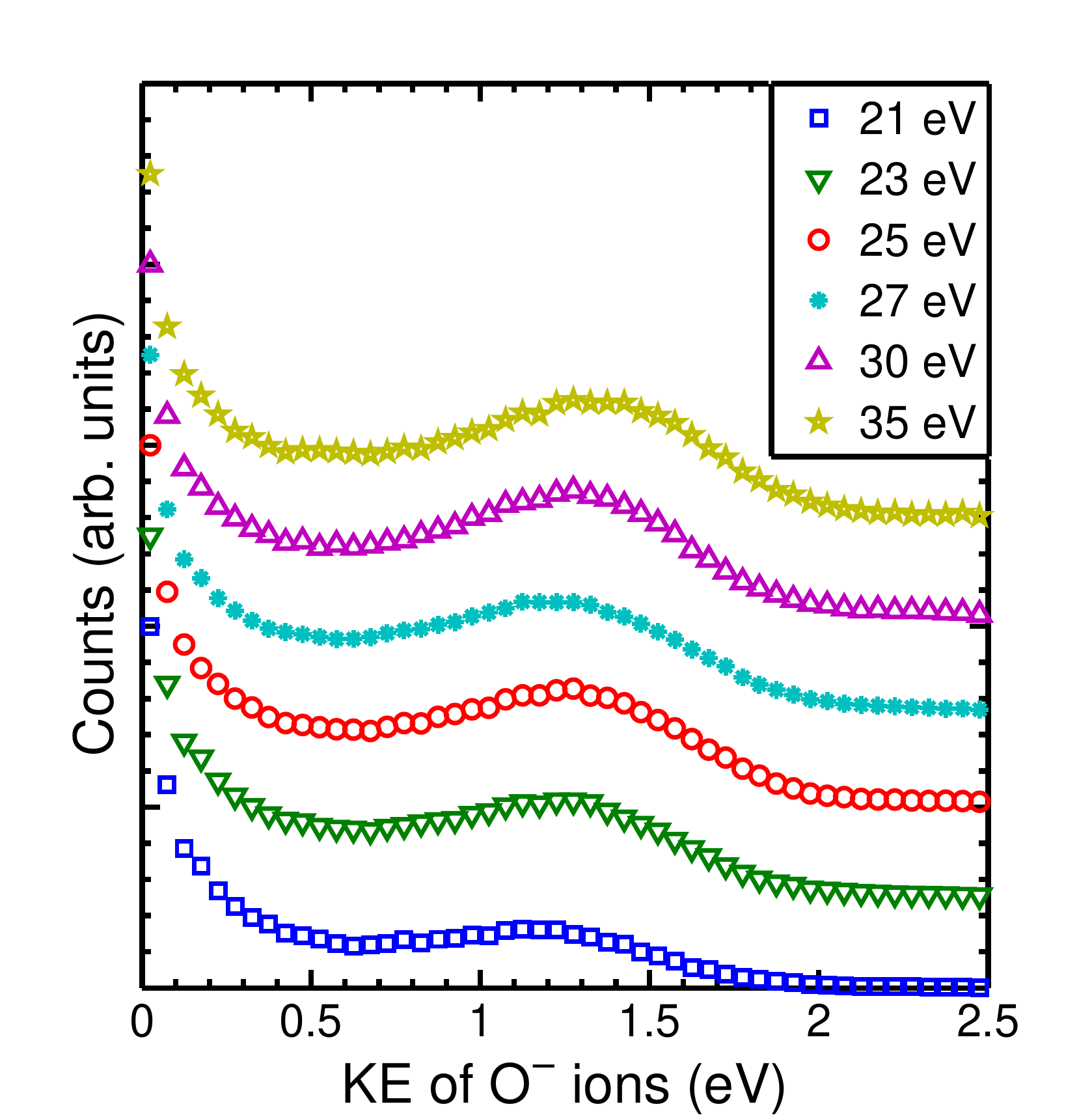}
\caption{The kinetic energy distribution of the O$^-$ ions produced from O$_2$ due dipolar dissociation for six different incident electron energies as indicated.} \label{fig:KE}
\end{center}
\end{figure}

The parameter $b$ is the constant background and $E_{Th}$ is the threshold of the DD process. The parameter $a$ is the scaling factor and set to zero below the threshold value and $n$ is the exponent. The effect of finite energy resolution of the electron beam is not considered in the set of equation~\ref{eq:th_fit}. Due to the nearly 0.8 eV broad energy spread of the incident electron beam the the ion yield curve gets smooth near the threshold which will lead to underestimation of the true threshold value. To minimize this effect only the data points slightly above from the apparent threshold upto 23.8 eV are considered in the fitting process and a nearly straight slope is obtained. The straight line is extrapolated until it intersects with the line obtained due to constant background ($f(E)=b$). The intersection point of these two curve is considered as the threshold value. Due to the effect of contact potential also the energy might get shift a little. The best fitted curve is obtained using the parameter $a=44.27$, $b=42.91$, $n=1.435$ and shown using a solid line in Fig. \ref{fig:threshold}. The threshold value for dipolar dissociation process is experimentally found to be 15.3 eV. The mismatch between the thremochemally obtained value 17.25 and experimentally obtained 15.3 eV could be due to poor electron beam energy resolution and the effect of contact potential.

The velocity slice images (VSI) of O$^-$ ions produced due to dipolar dissociation (DD) process for 21, 23, 25, 27, 30 and 35 eV energy electron collision with O$_2$ molecules are shown in Fig.~\ref{fig:VSI}. Out of the entire `Newton Sphere' of O$^-$ ions with nearly 500 ns time width a 50 ns thin flat slice through the central part has been taken for the kinetic energy and angular distribution measurement purpose. The incident electron beam direction is from left to right through the center of each images as indicated using small arrows on each of the VSI. 

\begin{figure}
\begin{center}
\includegraphics[scale=.4]{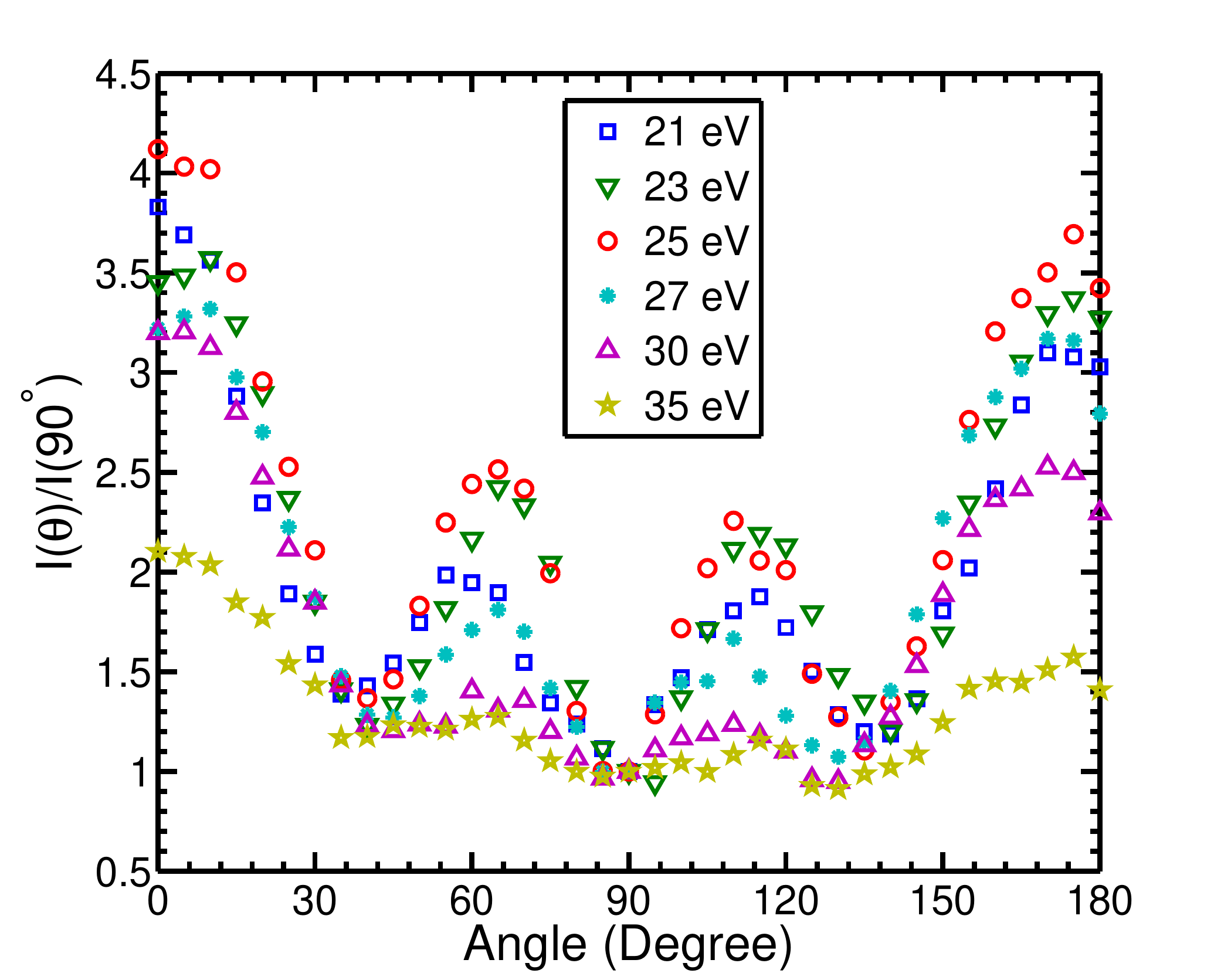}
\caption{Angular distribution of the O$^-$ ions created with kinetic energy in between 0.7 to 1.7 eV energy due to dipolar dissociation for six different incident electron energies.} \label{fig:ang_all}
\end{center}
\end{figure}
	
The kinetic energy distribution of the ions created due to dipolar dissociation are shown in Fig.~\ref{fig:KE}. The distributions are normalized at the near zero eV peak. The distributions show one large peak near zero eV followed by a broad peak in between 1.25 to 1.75 eV. The centre of the broad peak shifts towards right with increasing incident electron energy. A flat time slicing method is used to obtain the central slices containing the kinetic energy distribution information. So, in the kinetic energy distribution data only a fraction of the ions produced with higher kinetic energy are considered but all the ions with lower kinetic energy are included \cite{co2:moradmand2,inst:moradmand}. Thus the relative contributions of the ions with near zero eV and higher kinetic energy shown in Fig.~\ref{fig:KE} are not absolute. In the previous study of kinetic energy distribution by Van Brunt and Kieffer \cite{o2:polar_van1} a peak near 2 eV and another peak near 3.3 eV were reported. To best of our knowledge this is the only published kinetic energy distribution report of dipolar dissociation to O$_2$ molecules due to electron collision. The ions below 0.75 eV were not considered in that reported due to non-reliability. The higher peak near 3.3 eV has not been observed in the current study. In a recent study Nandi \emph{et al.} \cite{o2:polar_DN} also reported the absence of the higher peak from the VSIs. The near zero eV kinetic energy ions might not be created from a true dipolar states but due to pre-dissociation of the Rydberg states. The pre-dissociation is present throughout the entire dipolar dissociation region and ions with all possible kinetic energies in between 0 to around 2 eV can be observed. Pre-dissociation can occur if the life-time of an excited states is of the order of or more than few vibrational time period and the excited states cross the ion-pair states or interact with the vibrational continuum of ion-pair states. The ions with higher kinetic energy might be coming from excitation to true ion-pair states. If the ions created due to direct excitation to the ion-pair states have kinetic energy in between 1.25 to 1.75 eV then from equation~\ref{eq:th_fit} the energy transfer in the process comes out to be in between 19.75 to 20.75 eV. So, in the Franck-Condon transition region the separation between the ground state of oxygen molecule and the ion-pair state should be of the order of 19 eV. In ion-pair dissociation study of oxygen molecule using femtosecond depletion method Baklanov \emph{et al.} \cite{o2:baklanov} also found the ion-pair state around the same energy range.

\begin{figure*}
\begin{center}
\includegraphics[scale=.3]{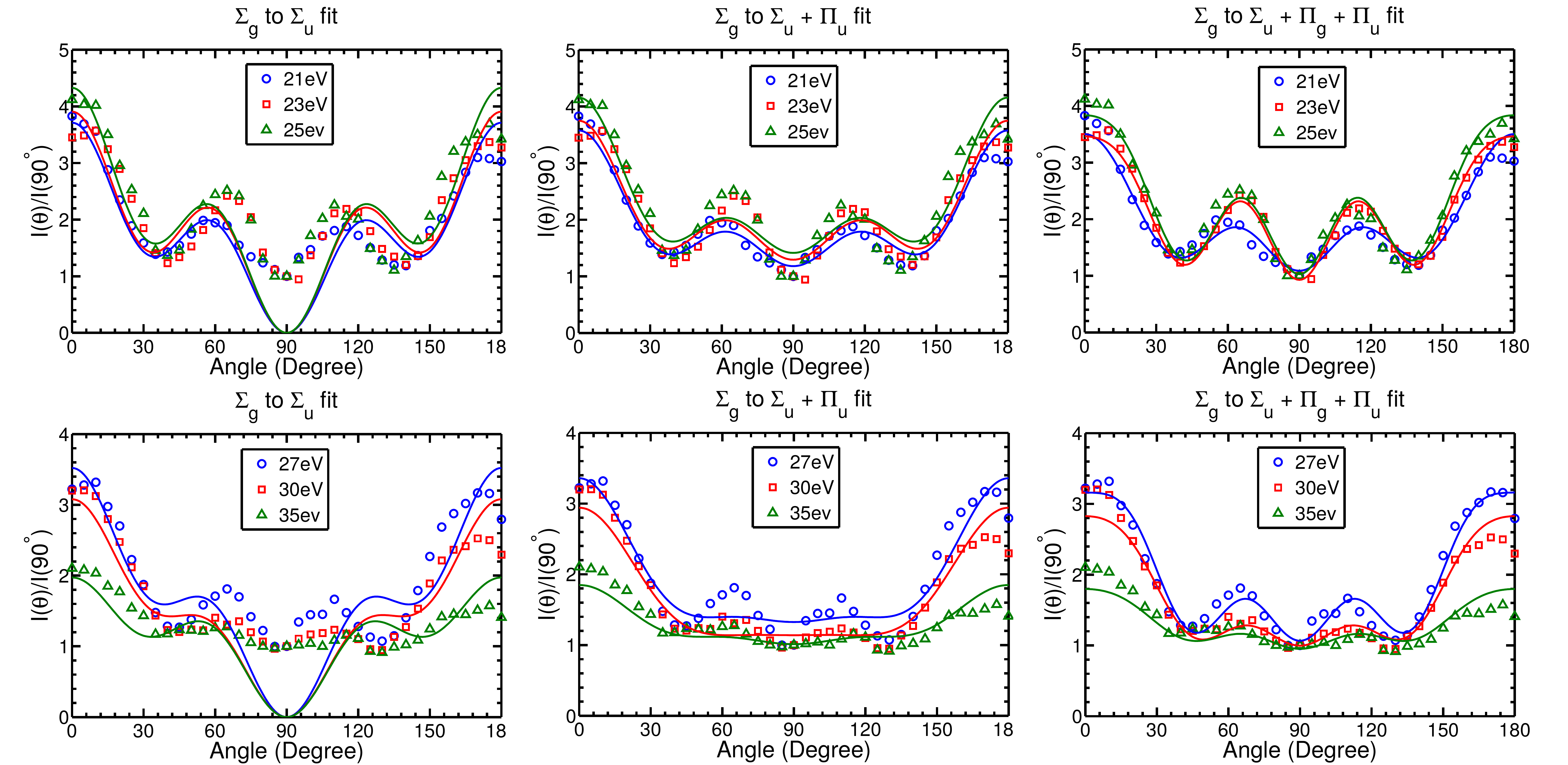}
\caption{The angular distribuition data for different incident electron energies, as indicated, are shown using the scattered points. In the left most column the solid lines indicate the fitted curve for a $\Sigma_g\rightarrow\Sigma_u$ transition. A transition from $\Sigma_g$ to $\Sigma_u+\Pi_u$ two states is shown using the solid curves in the middle column. The right most column shows the fitted curve for a $\Sigma_u, \Pi_g$ and $\Pi_u$ three final states transition process.} \label{fig:ang_fit}
\end{center}
\end{figure*}

The symmetry of the ion-pair states involved in the dipolar dissociation process can be determined from the angular distribution of the fragment ions. Due to energy and momentum conservation both the O$^-$ and O$^+$ ions formed in dipolar dissociation process will have similar kinetic energy and angular distribution. The angular distribution of fragment negative ions for six different incident electron energies in between 21 to 35 eV are shown in Fig.~\ref{fig:ang_all}. The ions created with kinetic energy in between 0.7 to 1.7 eV are only considered in angular distribution measurements. As already discussed due to flat slicing contribution from the entire `Newton Sphere' is present for the near zero eV ions thus no meaningful angular distribution information can be obtained for these ions from the sliced images  and are not discussed further. The angular distribution data are normalized at 90$^\circ$ and show two strong peaks at 0$^\circ$ and 180$^\circ$ and two small peaks around 65$^\circ$ and 115$^\circ$. For 21, 23 and 25 eV incident electron energies the anisotropy is prominent and increases with increasing energy. But, the anisotropic nature of the angular distribution starts decreasing after 27 eV incident electron energy and becomes almost isotropic with only a forward and backward peak. A forward-backward asymmetry observed in the angular distribution data might be caused by the effect of molecular recoil due electron collision \cite{ang:van}. 

The angular distribution of the fragments can be described using the expression given by Van Brunt \cite{o2:polar_van2} and identical with the expression derived by Zare \cite{zare} for the involvement of a single final state as

\begin{equation}
I(\Theta)\simeq K^{-n}\left|\sum_{l=\vert\mu\vert} ^\infty i^l\sqrt{\frac{(2l+1)(l-\mu)!}{(l+\mu)!}}\times j_l(\kappa)Y_{l,\mu}(\Theta,\Phi)\right|^2  \label{eq:ang1}
\end{equation}
Where, $\kappa$ is the product of momentum transfer vector between the incident and scattered electron, $K$, and the closest approach between the impinging electron and the center-of-mass of the molecule. The $j_l$'s are spherical Bessel function and $Y_{l,\mu}$'s are spherical harmonics. If $\Lambda_i$ and $\Lambda_f$ are the projection of electronic axial orbital angular momentum on the molecular axis for the initial and the final states respectively then, $\mu = \vert \Lambda_f - \Lambda_i \vert$. The angular momentum quantum number $l\geq \vert \mu \vert$ and restricted to have only even or odd values depending on whether the initial and final molecular states are having same of opposite parity for homonuclear molecules. In the present article the angular distribution is fitted using expression~\ref{eq:ang2}.

\begin{equation}
I\left(\theta\right)=\sum_{|\mu|}\left|\sum_{l=\left|\mu\right|}a_li^l\sqrt[]{\frac{\left(2l+1\right)\left(l-\mu\right)!}{\left(l+\mu\right)!}}j_l\left(\kappa\right)Y_{l,\mu}\left(\theta,\phi\right)e^{i\delta_l}\right|^2 \label{eq:ang2}
\end{equation}
The summation over $\mu$ takes care of the involvement of more than one final states. The phase differences between different partial waves involved in the transition with respect to the lowest one for each final state are expressed by $\delta_l$s. The momentum transfer vector $K$ and the adjustable parameter $n$ remain same for a particular incident electron energy and the term $K^{-n}$ in the expression~\ref{eq:ang1} can be absorbed within the parameter $a_l$ of the expression~\ref{eq:ang2}.

The ground state of O$_2$ molecule is $^3\Sigma_g ^-$, having $\Lambda_i= 0$. For the present case $\mu = 0$ and $1$ represents a transition from $\Sigma_g$ to $\Sigma$ and $\Pi$ state and even and odd values of $l$ are responsible for transition to a final state having garade ($g$) and ungarade ($u$) parity. To know about the symmetry of the ion-pair states involved in the process the angular distribution data has been fitted using expression~\ref{eq:ang2} and the fitting parameters are shown in Table~\ref{tab:tab1} and \ref{tab:tab2}. The angular distribution fitted with a $\Sigma_g$ to $\Sigma_u$ transition model is shown in the left most column of Fig.~\ref{fig:ang_fit}. This transition model can represents the angular distribution data upto a limit for 21 to 25 eV but with increasing incident electron energy the fit becomes poorer. This model also underestimate the value at 90$^\circ$. For a $\Sigma_u$ and $\Pi_u$ two final states transition model also the goodness of fitting decreases with increasing incident electron energy and shown in the middle column of Fig.~\ref{fig:ang_fit}. A transition from $\Sigma_g$ initial state to $\Sigma_u, \Pi_u$ and $\Pi_g$ three state transition model gives the best fit and shown in the right column of Fig.~\ref{fig:ang_fit}.  So, from the angular distribution data it is evident that a $\Sigma_u$ and $\Pi_u$ ion-pair state is present through out the entire incident electron energy range. A $\Pi_g$ state is also getting involved in the ion-pair production process and the contribution of this state increases with increasing incident electron energy and becomes significance above 27 eV incident electron energy. Van Brunt and Kieffer \cite{o2:polar_van1} also concluded that a $\Sigma_u$ state is responsible for the dipolar dissociation. In the limited angular distribution data the authors also found the presence of a $\Pi_u$ state. The molecules might get excited to a Rydberg state and then the Rydberg state can interact with an ion-pair state and dissociate into a O$^+$ and O$^-$ ions. Both the Rydberg and ion-pair state will have the same symmetry and in ion-pair dissociation study using XUV laser Zhou and Mo \cite{o2:zhou} found the symmetry of the excited states to be $\Pi_u$ and $\Sigma_u$. Dehmer and Chupka \cite{o2:chupka} measured the ion-pair production in 720 {\AA} to 670 {\AA} range (17.2 to 18.505 eV) and found the Rydberg states might dissociates via pre-dissociation through ion pair-states having symmetry $^3\Sigma_u ^-$ and $^3\Pi_u$. The authors also mentioned the dissociation is primarily through predissociation of the Rydberg states rather than by direct dissociation. But in the present case above 20 eV we believe both the predissociation and direct dissociation of ion-pair states are present and the higher energy ions considered in the angular distribution measurements are form direct dissociation of the ion-pair states.

\section{Conclusion}
The O$^-$ ions produced due to dipolar dissociation of O$_2$ molecule via interaction with electrons have been studied using time sliced velocity map imaging technique. From the kinetic energy distribution curve the location of the ion-pair state was found to be around  20 eV above the ground state in the Franck-Codon transition region. Pre-dissociation of a Rydberg state through an ion-pair state was found to be responsible for the dipolar dissociation process for relatively lower primary beam energy. Direct excitation to the ion-pair states was also found to be involved in dipolar dissociation process for relatively higher incident electron energies. From the angular distribution data a $\Sigma_u$ and a $\Pi_u$ ion-pair state was found to be present in the entire energy range. The involvement of a $\Sigma_g$ state is also observed with increasing incident electron energies. To know about the exact location of the ion-pair states in the potential energy surface and detail dynamics excitation to a ion-pair state either via pre-dissociation or direct excitation theoretical calculations are highly demanding.

\section{Acknowledgements}

D. N. gratefully acknowledges the partial financial support from ``Indian National Science Academy" for the development of VSI spectrometer under INSA Young Scientist project ``SP/YSP/80/2013/734".

\section*{References}
\bibliography{o2polar}

\begin{table*}
\centering
\caption{Fitting parameters for the angular distribution of the O$^{-}$ ions created due to 21, 23 and 25 eV incident electron energies for a \small{$\Sigma_g$ to $\Sigma_u$, $\Pi_g$ and $\Pi_u$ transition.}}\label{tab:tab1}
\begin{tabular}{c c c c c}
\hline
\hline
 & 21 eV & 23 eV & 25 eV\\
 \hline
\small{Weighting ratio of different partial waves} & & &\\
\small{$b_1$: $b_3$:}  & \small{1: 0.181:} & \small{1: 0.379:} & \small{1: 0.325:} \\
\small{$c_2$: $c_4$:}  & \small{1.359: 0.733:} & \small{0.048: 0.186:} & \small{0.076: 0.402} \\
\small{$e_1$: $e_3$} & \small{0.621: 0.786}  & \small{0.532: 0.438} & \small{0.572: 0.089}  \vspace*{.3cm}\\ 
\small{Phase difference} & & &\\
\small{$\delta_{\Sigma_u}, \delta_{\Pi_g}, \delta_{\Pi_u}$ (rad)} & \small{0.595, 0.426, 0.530 } & \small{0.726, 0.553, 0.330 } & \small{0.901, 0.812, 1.045}\\ 
\vspace*{.3cm}\\ 
\small{Parameter $\kappa_1, \kappa_2, \kappa_3 $} & 1.201, 2.204, 1.333 & 1.873, 1.702, 1.515 & 1.8, 1.985, 0.928\\
\hline
\hline
\end{tabular}
\end{table*}

\begin{table*}
\centering
\caption{Fitting parameters for the angular distribution of the O$^{-}$ ions created due to 27, 30 and 35 eV incident electron energies for a \small{$\Sigma_g$ to $\Sigma_u$, $\Pi_g$ and $\Pi_u$ transition.}} \label{tab:tab2}
\begin{tabular}{c c c c c}
\hline
\hline
 & 27 eV & 30 eV & 35 eV\\
 \hline
\small{Weighting ratio of different partial waves} & & &\\
\small{$b_1$: $b_3$:}  & \small{1: 0.832:} & \small{1: 0.246:} & \small{1: 0.187:} \\
\small{$c_2$: $c_4$:}  & \small{0.144: 0.032:} & \small{0.009: 0.981} & \small{0.081: 0.425} \\
\small{$e_1$: $e_3$} & \small{0.415: 0.333}  & \small{0.251: 0.064} & \small{0.708: 10.94}  \vspace*{.3cm}\\ 
\small{Phase difference} & & &\\
\small{$\delta_{\Sigma_u}, \delta_{\Pi_g}, \delta_{\Pi_u}$ (rad)} & \small{0.497, 0.847, 0.647 } & \small{0.606, 1.949, 0.055 } & \small{0.195, 0.490, 0.326}\\ 
\vspace*{.3cm}\\ 
\small{Parameter $\kappa_1, \kappa_2, \kappa_3 $} & 2.043, 1.042, 1.079 & 1.259, 2.108, 0.544 & 1.270, 2.025, 1.469 \\
\hline
\hline
\end{tabular}
\end{table*}
\end{document}